\documentstyle[floats,twocolumn,aps,prl] {revtex}

\def\compoundrel#1\over#2{\mathpalette\compoundreL{{#1}\over{#2}}}
\def\compoundreL#1#2{\compoundREL#1#2}
\def\compoundREL#1#2\over#3{\mathrel
     {\vcenter{\hbox{$\buildrel{#1#2}\over{#1#3}$}}}}

\begin{document}

\title{Scattering of hydrogen molecules from a reactive surface:\\
Strong off-specular and rotationally inelastic diffraction} 

\author{Axel Gross and Matthias Scheffler}

\address{Fritz-Haber-Institut der Max-Planck-Gesellschaft, Faradayweg 4-6, 
D-14195 Berlin-Dahlem, Germany}

\twocolumn[
\maketitle

\vspace{-.7cm}

\begin{abstract}

\hspace*{1.5cm}\parbox{15cm}{
Six-dimensional quantum dynamical calculations of the scattering 
of H$_2$ from a Pd\,(100) surface using a potential energy surface 
derived from density-functional theory calculations are presented.
Due to the  corrugation and anisotropy of the PES
strong off-specular and rotationally inelastic diffraction 
is found. The dependence of the diffraction intensitities
on the incident kinetic energy is closely examined.
In particular we focus on the quantum oscillations
for normal and off-normal incidence.
}
\end{abstract}

\hspace*{1.5cm}\parbox{15cm}{
}

\vspace{-.4cm}

]

Coherent elastic scattering of atoms or molecules from surfaces is a tool for 
probing surface structures and adsorbate--substrate interaction potentials. In
particular helium atom scattering (HAS) has been used intensively to study
surface crystallography and the shape of physisorption potentials (see, e.g.,
\cite{Hulpke} and references therein). Hydrogen molecules
has been utilized less frequently in order to study interaction potentials. 
The coherent elastic scattering of molecules is more complex 
than atom scattering 
because in addition to parallel momentum transfer
the internal states of the molecule can be changed in scattering.
This means in particular rotational transitions, because 
the energies typically employed in molecular scattering experiments
are not sufficient to excite molecular vibrations.

Rotational transitions in elastic coherent scattering
lead to the appearance of additional peaks in the
diffraction pattern since the total kinetic energy of 
the molecule is changed. Most experimental studies of rotationally
inelastic scattering of hydrogen have been performed at the surfaces
of ionic solids \cite{Row75,Bru83}. (Here ``inelastic'' refers to
energy exchange between the hydrogen degrees of freedom, not with
respect to the substrate.)
In the hitherto existing experiments of hydrogen scattering from metal 
surfaces the additional peaks could usually hardly be resolved
\cite{Lap81,Rob85,Wha85} except in the case of HD scattering 
\cite{Wha85,Cow83,Ber90}, where the displacement of the center of mass 
from the center of the charge distribution leads to a strong rotational
anisotropy. Only recently Bertino {\it et al.} \cite{Ber96} 
clearly resolved rotationally inelastic peaks in the
diffraction pattern of D$_2$/Ni(110) in addition to rotationally elastic 
peaks, which had been observed before by Robota {\it et al.} \cite{Rob85}. 
Furthermore, Cvetko {\it et al.} \cite{Cve96}
reported diffractive and rotationally inelastic scattering of D$_2$/Rh(110).

In the case of hydrogen scattering from
reactive surfaces, the repulsive interaction is
not mediated by the tail of the metal electron density, but occurs rather
close to the surface \cite{Wil95,Kay95}. Due to the chemical nature
of this interaction the potential is strongly corrugated and anisotropic
with regard to the molecular orientation. This requires a realistic 
microscopic description of the scattering dynamics.
Existing theoretical studies of hydrogen scattering at surfaces have
either treated non-reactive systems \cite{Wol75,Gar76,Lev89,Kro95a,Kro95b}
or they were restricted to low dimensions on model 
potentials \cite{Hal88,Dar90,Dar92}.

We have performed six-dimensional quantum dynamical calculations
of the scattering of H$_2$ from Pd/(100) on a PES that was derived
from density-functional theory calculations~\cite{Wil95,Gro95}. 
The PES exhibits
a coexistence of activated with non-activated paths to adsorption 
together with a broad distribution of barrier heights.
{\em All} six degrees of freedom of the hydrogen molecule are treated 
quantum dynamically in a coupled-channel scheme \cite{Bre93,Bre94} while 
the substrate atoms are kept fixed; for computational details we
refer to Ref.~\cite{Gro96d}.

\vspace{-12.1cm}

\hspace*{-1.0cm} {\large {\tt subm to Chem.~Phys.~Lett., July 1996}}

\vspace{11.7cm}

Our results show
strong off-specular and rotationally inelastic diffraction 
which are caused by the  corrugation and anisotropy of the PES.
The intensities of the diffraction peaks exhibit a pronounced
oscillatory structure as a function of the incident energy. 
We have already predicted such an structure for the sticking
probability of H$_2$/Pd\,(100) \cite{Gro95}. Very recently Rettner and
Auerbach searched for these oscillations experimentally \cite{Ret96},
but did not find any. We will carefully discuss the physical origin
of the oscillations and point out why they might not have been 
detected. Hence this paper may also be regarded as an extended 
comment to Ref.~\cite{Ret96}.

\begin{figure}[tb]
\unitlength1cm
\begin{center}
   \begin{picture}(10,6.5)
      \includegraphics{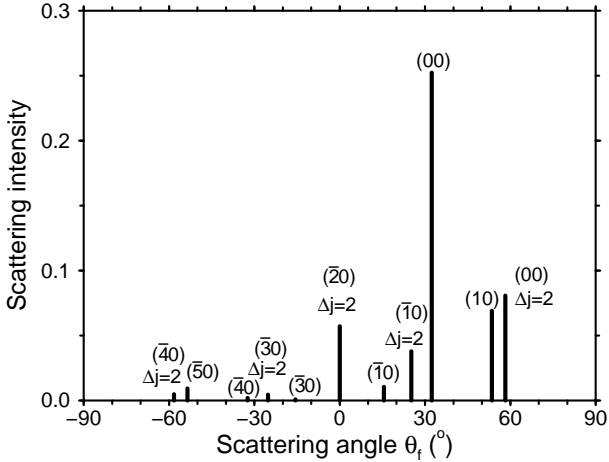}
   \end{picture}

\end{center}
   \caption{Angular distribution of scattered H$_2$ molecules.
            The initial kinetic energy is $E_i = 76$~meV, the
            incident angle is $\theta_i = 32^o$ along the
            $\langle0 \bar 1 1\rangle$ direction. The molecules are initially
            in the rotational ground state $j_i = 0$.
            The rotationally inelastic diffraction peaks have
            been marked with $\Delta j =2$.   }

\label{scatter}
\end{figure}

One typical calculated angular distribution of H$_2$ molecules 
scattered at Pd(100) is shown in Fig.~\ref{scatter}. The total
initial kinetic energy is $E_i = 76$~meV.
The incident parallel momentum corresponds to $2\hbar G$
along the $\langle0 \bar 1 1\rangle$ direction, where
$G= 2\pi/a$, $a$ is the nearest neighbor distance between two Pd atoms.  
This leads to an incident angle of $\theta_i = 32^{\circ}$. The molecules
are initially in the rotational ground state $j_i = 0$. 
$(m,n)$ denotes the parallel momentum transfer
$\Delta {\bf G}_{\|} = (mG,nG)$. 
The specular peak is the most pronounced one, but the first 
order diffraction peak (10) is only a factor of four
smaller (note, that in a typical HAS experiment the off-specular
peaks are about two orders of magnitude smaller than the specular
peak). The results for the rotationally inelastic diffraction
peaks $j = 0 \rightarrow 2$ have been summed over all final
azimuthal quantum numbers $m_j$, as has been done for all
results presented in the following.
 The excitation probability of the so-called
cartwheel rotation with $m = 0$ is for all peaks approximately one
order of magnitude larger than for the so-called helicopter
rotation $m = j$, since the polar anisotropy of the PES is stronger 
than the azimuthal one.

This steric effect in scattering could already be expected from detailed
balance arguments. According to the principle of detailed balance,
in a equilibrium situation the flux impinging on a surface from the
gas-phase, which is rotationally isotropically distributed, should equal
the desorption plus the scattering flux. Since in desorption the
helicopter rotations are preferentially occupied \cite{Gro95,Wet96},
in scattering the cartwheel rotations have to be preferentially
excited. 

The intensity of the rotationally inelastic diffraction peaks 
in Fig.~\ref{scatter} is comparable to the rotationally elastic ones.
Except for the specular peak they are even larger than
the corresponding rotationally elastic diffraction peak with
the same momentum transfer $(m,n)$. Note that due to the initial
conditions the rotationally elastic and inelastic $({\bar 2} 0)$
diffraction peaks fall upon each other.

\begin{figure}[tb]
\unitlength1cm
\begin{center}
   \begin{picture}(10,6.1)
      \includegraphics{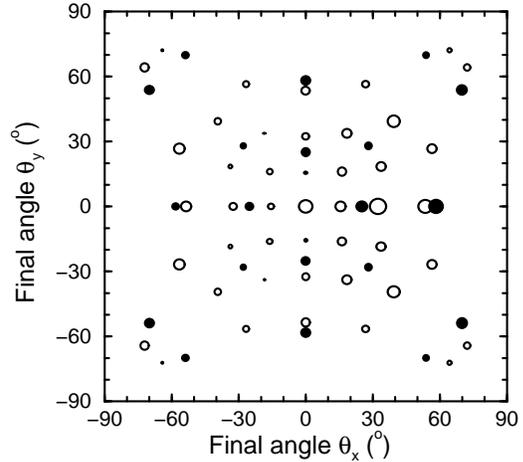}
   \end{picture}

\end{center}
   \caption{Angular distribution of the in-plane and out-of-plane
           of scattering of H$_2$/Pd(100). The initial conditions 
           are the same as in Fig.~\protect{\ref{scatter}}.
           Open circles correspond to rotationally elastic, filled
           circles to rotationally inelastic diffraction.
           The radii of the circles are proportional to the logarithm 
           of the scattering intensity.  
           $x$ denotes the $\langle0 \bar 1 1\rangle$ direction, $y$ the
           $\langle0 1 1\rangle$ direction. The specular peak is the largest
           open circle. }

\label{inout}
\end{figure}

The out-of-plane scattering intensities are not negligible, which is
demonstrated in Fig.~\ref{inout}. The initial conditions are the same
as in Fig.~\ref{scatter}. The open circles correspond to rotationally
elastic, the filled circles to rotationally inelastic diffraction.
The radii of the circles are proportional to the logarithm of the
scattering intensity. The sum of all out-of-plane scattering intensities
is approximately equal to the sum of all in-plane scattering intensities. 
Interestingly, some diffraction peaks with a large parallel
momentum transfer still show substantial intensities. This phenomenon
is well known form helium atom scattering and has been discussed
within the concept of so-called rainbow scattering \cite{Gar75}.

\begin{figure}[tb]
\unitlength1cm
\begin{center}

   \begin{picture}(10,11)
      \includegraphics{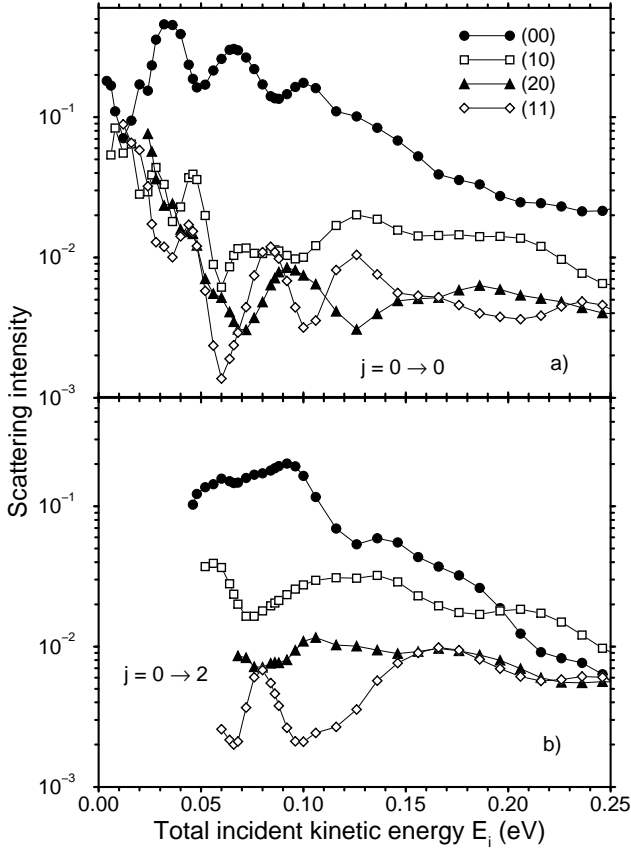}
   \end{picture}

\end{center}
   \caption{Scattering intensity versus kinetic energy 
            with an initial velocity spread $\Delta v_{i}/v_i = 0.05$ 
            for H$_2$ molecules in the rotational ground state
            under normal incidence. a) Rotationally elastic 
            $j = 0 \rightarrow 0$ scattering,
            b) rotationally inelastic $j = 0 \rightarrow 2$ scattering. 
            The intensities are summed over all final
            azimuthal quantum number $m_j$. }

\label{diffper}
\end{figure} 

\begin{figure}[tb]
\unitlength1cm
\begin{center}

   \begin{picture}(10,12)
      \includegraphics{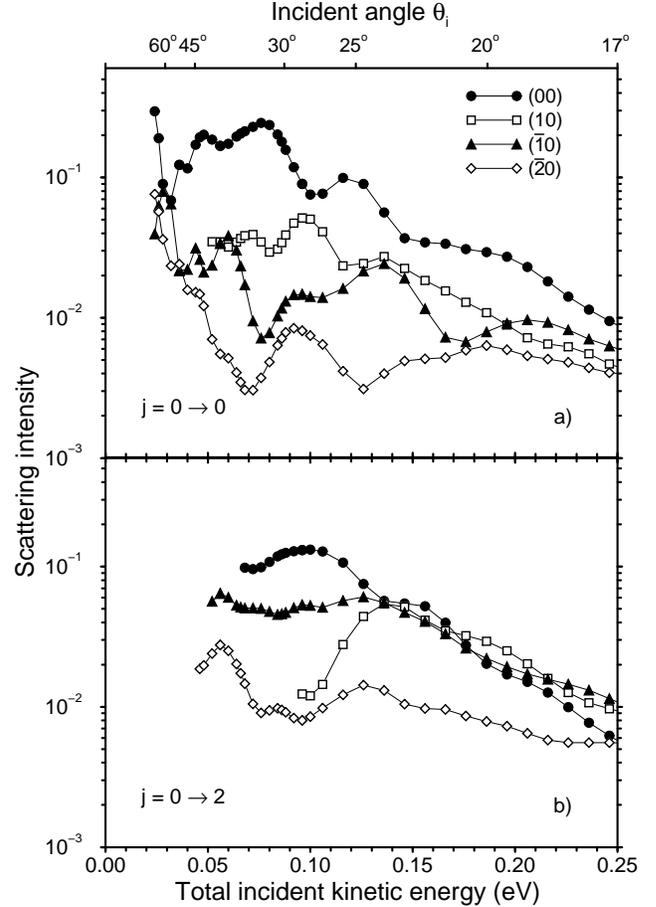}
   \end{picture}

\end{center}
   \caption{Scattering intensity versus total kinetic energy 
            with an initial velocity spread $\Delta v_{i}/v_i = 0.05$ 
            for H$_2$ molecules in the rotational ground state
            under off-normal incidence. The incident parallel momentum
            corresponds to $2\hbar G$ along the $\langle0 \bar 1 1\rangle$ 
            direction with $G = 2\pi/a$. The incident
            angle is given at the top axis.
            a) Rotationally elastic 
            $j = 0 \rightarrow 0$ scattering,
            b) rotationally inelastic $j = 0 \rightarrow 2$ scattering. 
            The intensities are summed over all final
            azimuthal quantum number $m_j$. }

\label{diff2G}
\end{figure}

\begin{figure}[tb]
\unitlength1cm
\begin{center}
   \begin{picture}(10,6.5)
      \includegraphics{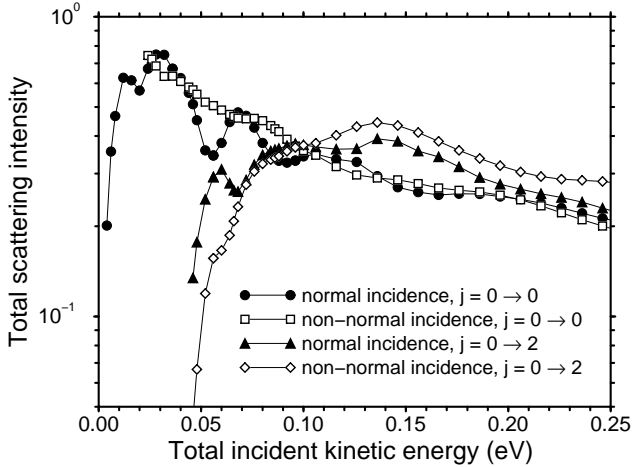}
   \end{picture}

\end{center}
   \caption{Total scattering intensity versus total kinetic energy
            with an initial velocity spread $\Delta v_{i}/v_i = 0.05$ 
            for rotationally elastic and inelastic scattering at
            normal incidence and off-normal incidence with initial 
            parallel momentum $p_{\parallel} = 2\hbar G$.   }

\label{difftot}
\end{figure}

A more detailed analysis of the diffraction intensities versus
total incident kinetic energy is shown in Figs.~\ref{diffper}-\ref{difftot}.
In Fig.~\ref{diffper} the intensities of four different diffraction peaks
are plotted for normal incidence; Fig.~\ref{diffper}a) shows rotationally
elastic, Fig.~\ref{diffper}b) rotationally inelastic diffraction. 
In the experiment the molecular beams are not monoenergetic but have
a certain velocity spread. In order to make close contact to experiments
we assumed for our results in Figs.~\ref{diffper}-\ref{difftot} an initial 
velocity spread $\Delta v_{i}/v_i = 0.05$, close to the  
velocity spread of recent experiments~\cite{Ber96}.

The theoretical data still exhibit a rather strong oscillatory
structure which is a consequence of the quantum nature of hydrogen
scattering. We first focus on the specular peak in Fig.~\ref{diffper}a).
The oscillations at low kinetic energies reflect the opening up
of new scattering channels \cite{Gro95b}. The first pronounced dip
at $E_i = 12$~meV coincides with the opening up of the (11)~diffraction
peak, the small dip at $E_i = 22$~meV with the opening up of the 
(20)~diffraction channel. The huge dip at approximately 50~meV
reflects the threshold for rotationally inelastic scattering.
Interestingly, the rotational elastic (10) and (11) diffraction 
intensities show pronounced maxima at this energy. This indicates
a strong coupling between parallel motion and rotational excitation.

At larger energies the scattering intensities still show an
oscillatory structure. In addition to the opening up of new scattering
channels this also reflects the existence of scattering resonances
\cite{Dar90}: molecules can become temporarily trapped into 
metastable molecular adsorption states at the surface due to the
transfer of normal momentum to parallel and angular momentum
which resonantly enhances the scattering intensity. 

These structures are known for a long time
in He and H$_2$ scattering \cite{stern} and also in LEED \cite{mcrae,pendry}. 
For H$_2$/Pd(100), however, measuring these oscillations is a very demanding
task. They are very sensitive to surface imperfections like adatoms or steps.
Due to the high reactivity of this system either an adlayer of hydrogen atoms
builds up very rapidly during the experiment or in order to avoid this
one has to go to high surface temperatures \cite{Gro96c}. Both
suppresses the oscillations.

Figure~\ref{diffper}b) shows the intensities of rotationally inelastic
diffraction peaks. The specular peak is still the largest one, 
however, some off-specular peaks become larger
than the (00) peak at higher energies. This is due to the fact that the 
rotationally anisotropic component of the potential is more corrugated than
the isotropic component \cite{Wil95}. Furthermore it is apparent that
the oscillatory structure is somewhat smaller for rotationally
inelastic than for rotationally elastic scattering.

The results shown in Fig.~\ref{diff2G} correspond to a H$_2$ beam
impinging on the Pd(100) surface under off-normal incindence. 
The initial parallel momentum is $p_{\parallel} = 2 \hbar G$ 
along the $\langle0 \bar 1 1\rangle$ direction. Note that the incident angle
decreases with increasing total kinetic energy for constant
initial parallel momentum.
Only in-plane scattering intensities are plotted. 
Now the intensities of the off-specular peaks compared to the specular
peak are much larger than in the case of normal incidence, which means
that the change of parallel momentum in molecule-surface scattering is
much more probable for off-normal than for normal incidence. 
Again a pronounced oscillatory structure for the single
diffraction peaks is observed.

One overall trend is evident in Fig.~\ref{diffper} and \ref{diff2G}: 
the scattering intensities of almost all single peaks decrease with
increasing kinetic energy as also observed in the experiment \cite{Ber96}, 
whereby the specular peak decreases more strongly
than the off-specular peaks. 
In ref.~\cite{Ber96} the energetic dependence of the scattering intensities
is discussed within a Debye-Waller analysis in the impulsive
and harmonic approximation \cite{Hulpke}; the decrease in the peak
height with increasing energy is solely attributed to the interaction
with surface phonons. Our analysis, like others before 
\cite{Hal88,Dar90,Dar92}, shows that there is already a substantial
decrease in the intensities for hydrogen scattering at metal surfaces
without taking any energy transfer to
the phonons into account.

In Fig.~\ref{difftot} we have plotted the total reflection probability 
for rotationally elastic and rotationally inelastic $j = 0 \rightarrow 2$ 
scattering, i.e. the sum over all corresponding diffraction peaks.  
After an initial increase, which corresponds to the decrease of the
sticking probability at low energies \cite{Gro95,Ren89}
,  the reflection probabilities
generally decrease with increasing energy. 
For kinetic energies above 0.1~eV the probability of rotationally
inelastic $j = 0 \rightarrow 2$ scattering becomes larger than the
probability of rotationally elastic scattering. In addition, the rotational
excitation is larger for off-normal incidence than for normal incidence.
The oscillatory structure of the total scattering intensities is less
pronounced (note the different intensity scale) than for the single
diffraction peaks, but is still clearly visible, especially the
threshold for rotational excitation at normal incidence.

Note that the minimum in the specular peak at \mbox{$E_i \approx 12$~meV} has
now turned into a local maximum in the total scattering intensity.
That is due to the fact that at this energy not only one scattering
channels opens up, but actually four due to symmetry, namely the
(10), ($\bar 1$0), (01) and (0$\bar 1$) peaks. The additional 
scattering intensity in these diffraction peaks over-compensates
for the loss of the specular peak in the total scattering intensities.

Although the scattering intensities of the single peak for off-normal 
incidence also showed a strong oscillatory structure, their sum 
is rather smooth (see Fig.~\ref{difftot}). This can also be understood by
symmetry arguments. For an incident angle along the 
$\langle0 \bar 1 1\rangle$ direction the (10), ($\bar 1$0), (01) and 
(0$\bar 1$) peaks, e.g., open up at three different energies, for a general
incident azimuth even at four different energies. Therefore the effect
of the opening up of new scattering channels is significantly reduced for 
the total scattering intensity compared to normal incidence.
As far as quantum oscillations are concerned, it is therefore not reasonable
to compare experiments for an angle of incidence of $\theta_i = 15^{\circ}$
\cite{Ret96} with calculations for normal incidence.

In order to observe quantum oscillations we suggest not to determine the 
whole scattering flux for off-normal incidence as done in ref.~\cite{Ret96},
but rather monitor the intensity of a single diffraction peak as a function
of the kinetic energy. Thus one also avoids to include incoherently 
scattered molecules due to impurity or multi-phonon scattering in the
measurements. In our calculations the substrate is held fixed. Substrate 
motion certainly makes the detection of quantum oscillations more difficult.
Since these oscillations have been observed for H$_2$ scattering at
other surfaces \cite{stern}, they should, however, also be detectable 
for H$_2$/Pd\,(100) scattering.

The general decrease of the total scattering intensities
at higher energies reflects the fact
that the sticking probability rises in this energy range so that
less molecules are reflected.
However, the decrease of the total intensities is much slower
than the decrease of the single peaks
in Figs.~\ref{diffper} and \ref{diff2G} (note again the different intensity
scale). This implies that the decrease in the intensities of the single peaks
is partly compensated by the opening up of new scattering channels with
increasing kinetic energy. The number of energetically accessible
parallel momentum states increases roughly linearly with energy, as does
the number of rotational states, so that the number of accessible
scatterings states increases quadratically with increasing energy
(again neglecting vibrational excitation). Hence the increase of the
available phase space for scattering with increasing kinetic energy
is responsible for the stronger decrease of the single diffraction
peaks compared to the total reflection probability.

In conclusion, we have performed a six-dimensional quantum dynamical 
study of the scattering of H$_2$ from a Pd\,(100) surface using a potential
energy surface (PES) derived from first-principles calculation.
The  corrugation and anisotropy of the PES lead to strong off-specular 
and rotationally inelastic diffraction intensities.
Strong quantum oscillations in the diffraction intensities as a
function of the kinetic energy are obtained. Especially at low kinetic
energies they reflect the opening up of new scattering channels
with increasing kinetic energy.

We like to thank M.F. Bertino and J.P. Toennies for sending us their
manuscript prior to publication. In addition, useful discussions
with M.F. Bertino are gratefully acknowledged.


\vspace{-.5cm}

\begin{thebibliography}{99}

\vspace{-1.7cm}

\bibitem{Hulpke} E. Hulpke, Ed., Helium Atom Scattering from Surfaces 
(Springer, Berlin, 1992).
\bibitem{Row75} R.G. Rowe and G. Ehrlich, J. Chem. Phys. {\bf 63}, 
4648 (1975).
\bibitem{Bru83} G. Brusdeylins and J.P. Toennies, Surf. Sci. {\bf 126},
647 (1983).

\bibitem{Lap81} J. Lapujoulade, Y. Le Cruer, M. Lefort, Y. Lejay, and
E. Maurel, Surf. Sci. {\bf 103}, L85 (1981)
\bibitem{Rob85} H.J. Robota, W. Vielhaber, M.C. Lin, J. Segner, and G. Ertl,
Surf. Sci. {\bf 155}, 101 (1985). 
\bibitem{Wha85} K.B. Whaley, C. Yu, C.S. Hogg, J.C. Light, and S.J Sibener,
J. Chem. Phys. {\bf 83}, 4325 (1985).

\bibitem{Cow83} J.P. Cowin, C. Yu, S.J Sibener, and L. Wharton, 
J. Chem. Phys. {\bf 79}, 3537 (1983).
\bibitem{Ber90} R. Berndt, J.P. Toennies, and. Ch. W{\"o}ll,
J. Chem. Phys. {\bf 92}, 1468 (1990).

\bibitem{Ber96} M.F. Bertino, F. Hofmann, and J.P. Toennies, J. Chem. Phys.,
in press. 
\bibitem{Cve96} D. Cvetko, A. Morgante, A. Santaniello and F. Tommasini,
J. Chem. Phys. {\bf 104}, 7778 (1996).

\bibitem{Wil95} S. Wilke and M. Scheffler, Surf. Sci. {\bf 329}, L605 (1995),
Phys. Rev. {\bf B 53}, 4926 (1996). 

\bibitem{Kay95} M. Kay, G.R. Darling, S. Holloway, J.A. White, and D.M. Bird,
Chem. Phys. Lett. {\bf 245}, 311 (1995).


\bibitem{Wol75} G. Wolken, J. Chem. Phys. {\bf 62}, 2730 (1975).
\bibitem{Gar76} U. Garibaldi, A.C. Levi, R. Spadacini, and G.E. Tommei,
Surf. Sci. {\bf 55}, 40 (1976).
\bibitem{Lev89} A.C. Levi and V. Tarditi, Surf. Sci. {\bf 219}, 235 (1989).
\bibitem{Kro95a} G.J. Kroes and R.C. Mowrey, Chem. Phys. Lett. {\bf 232}, 258
(1995). 
\bibitem{Kro95b} G.J. Kroes and R.C. Mowrey, J. Chem. Phys. {\bf 103}, 2186
(1995). 

\bibitem{Hal88} D. Halstead and S. Holloway, J. Chem. Phys {\bf 88}, 7197
(1988).
\bibitem{Dar90} G.R. Darling and S. Holloway, J. Chem. Phys {\bf 93}, 9145
(1990).
\bibitem{Dar92} G.R. Darling and S. Holloway, J. Chem. Phys {\bf 97}, 5182
(1992).



\bibitem{Gro95} A. Gross, S. Wilke, and M. Scheffler, Phys. Rev. Lett. 
{\bf 75}, 2718 (1995).


\bibitem{Bre93} W. Brenig, T. Brunner, A. Gross, and R. Russ, Z.~Phys.~B 
{\bf 93}, 91 (1993).
\bibitem{Bre94} W. Brenig and R. Russ, Surf. Sci. {\bf 315}, 195 (1994).

\bibitem{Gro96d} A. Gross and M. Scheffler, in preparation.


\bibitem{Ret96} C.T. Rettner and D.J. Auerbach, Chem. Phys. Lett. {\bf 253},
236 (1996).



\bibitem{Wet96} D. Wetzig, R. Dopheide, M. Rutkowski, R. David, and
H. Zacharias, Phys. Rev. Lett. {\bf 76}, 463 (1996).


\bibitem{Gar75} U. Garibaldi, A.C. Levi, R. Spadacini, and G.E. Tommei,
Surf. Sci. {\bf 48}, 649 (1975).

\bibitem{Gro95b} A. Gross, J. Chem. Phys. {\bf 102}, 5045 (1995).

\bibitem{stern}  R. Frisch and O. Stern, Z. Phys. {\bf 84}, 430 (1933).


\bibitem{mcrae} E. G. McRae, Rev. Mod. Phys. {\bf 51}, 541 (1979).

\bibitem{pendry}  J.B. Pendry, {\it Low energy electron diffraction}, Academic
Press, London (1974), p.112.

\bibitem{Gro96c} A. Gross and M. Scheffler, Phys. Rev. Lett. {\bf 77}, 
405 (1996).


\bibitem{Ren89} K. D. Rendulic, G. Anger, and A. Winkler, Surf. Sci. 
{\bf 208}, 404 (1989). 


\end{thebibliography}
\end{document}